# Beam Measurements of the Tianlai Dish Radio Telescope using an Unmanned Aerial Vehicle

Juyong Zhang (zhang_juyong01@163.com), Hangzhou Dianzi University

Jingxin Liu (ljx171010010@hdu.edu.cn), Hangzhou Dianzi University

Fengquan Wu (wufq@bao.ac.cn), National Astronomical Observatories, Chinese Academy of Science

Xuelei Chen (xuelei@bao.ac.cn), National Astronomical Observatories, Chinese Academy of Science

Jixia Li (jxli@bao.ac.cn), National Astronomical Observatories, Chinese Academy of Sciences

Peter T. Timbie (pttimbie@wisc.edu), Department of Physics at the University of Wisconsin – Madison

Santanu Das (sanjone@gmail.com), Department of Physics at the University of Wisconsin – Madison, Fermi National Accelerator Laboratory

Ruibin Yan (yrb@hdu.edu.cn), Hangzhou Dianzi University

Jiachen He (2252699h@student.gla.ac.uk), University of Glasgow

Osinga Calvin (calvinosinga@gmail.com), Department of Physics at the University of Wisconsin-Madison

Precision measurement of the beam pattern of an antenna is very important for many applications. While traditionally such measurement is often made in a microwave anechoic chamber or at a test range, measurement using an unmanned aerial vehicle (UAV) offers a number of advantages: the measurement can be made for the assembled antenna on site, thus reflecting the actual characteristics of the antenna of interest, and more importantly, it can be performed for larger antennas which cannot be steered or easily measured using the anechoic chamber and test range. Here we report our beam measurement experiment with UAV for a 6 meter dish used in the Tianlai array, which is a radio astronomy experiment. Due to the dish's small collecting area, calibration with an astronomical source only allows for determining the antenna beam pattern over a very limited angular range. We describe in detail the setup of the experiment, the components of the signal transmitting system, the design of the flight path and the procedure for data processing. We find the UAV measurement of the beam pattern agrees very well with the astronomical source measurement in the main lobe, but the UAV measurement can be extended to the fourth side lobe. The measured position and width of each lobe also shows good agreement with electromagnetic field simulation. This UAV-based approach of beam pattern measurement is flexible and inexpensive, and the technique may also be applied to other experiments.

Beam measurement, unmanned aerial vehicle, dish telescope

# 1 Introduction

The beam profile of an antenna can be measured by scanning it over a calibration source in the far field, and recording the output power from the antenna as a function of the angle between the direction of the source and the antenna axis. For large antennas which do not fit into an anechoic chamber, measurements are often made in a test range where the source is mounted at a large distance and height. However, this method has certain limitations, as constraints on the height of the source make measurements at high altitude angles difficult, and the measurement may also suffer from reflections from the ground. Moreover, the antennas thus measured still need to be shipped to their installation site and re-assembled; in this process the beam profile may change, so the accuracy of the measured profile is degraded. On-site calibration of the antenna beam may be carried out with sources in the field (e.g., a broadcasting artificial satellite [1] or a strong astronomical radio source [2]). However, the artificial satellite transmission is limited to certain specific frequencies, while even the strongest radio astronomical source has relatively low flux, so very often it can only provide calibration within the main lobe of the beam with limited precision.

The unmanned aerial vehicle or, informally, the "drone," provides an economical means to measure the antenna beam profile. It can carry a calibration source and fly over the antenna under test (AUT), allowing careful measurements to be made. Compared with a tethered balloon, it can be more easily maneuvered to and kept at the desired position for the measurement. Recently this technique has been employed by a number of research groups to calibrate antennas, especially radio telescopes used in astronomy. Chang et al. [3] used an UAV to calibrate a 5 m dish telescope at 1.2 GHz. The fourth side lobe of the AUT was measured successfully. In their experiment, the absolute positioning accuracy of the UAV was a few meters. The noise transmitter horn was held by a gimbal to ensure that it was pointing directly downward during the measurement. The UAV measurement compared well with other beam calibration methods. Picar et al. [4] used an UAV equipped with a monopole antenna to measure a 5 m dish telescope at 328 MHz. The positioning accuracy of the UAV used in their measurement was within a meter. The measurement results showed good agreement with the simulation. Jacobs et al. [5] utilized a UAV to calibrate the beam patterns of two identical dual polarization antennas used in the Orbcomm system at 137 MHz. The two AUTs were separated by 100 m and the flight path was a spherical shell of radius 100 m centered on each AUT. The measured beam patterns of the two AUTs show good agreement with each other and with the model. Acedo et al. [6] designed an UAV-based measurement system for the AAVS0 array, which was a test bed for the SKA1-LOW instrument. Positions of 16 elements in the AAVS0 array were measured by photogrammetry. Both a single antenna and an antenna array were measured at 50-350 MHz in their experiment. These works prove that the UAV-based approach is very useful for beam pattern measurement.

Here we present our calibration of a Tianlai dish telescope using the UAV. The Tianlai (Chinese meaning "heavenly sound") project is a 21cm intensity mapping pathfinder experiment in China [7–9], as shown in Figure 1. Its science objective is to observe the distribution of neutral hydrogen (HI) in the Universe, and from the HI distribution, to measure the baryon acoustic oscillation (BAO) signal in the large scale structure. The latter can be used as a standard ruler to determine the expansion rate of the universe at different redshifts and infer the dark energy equation of state parameter. The neutral hydrogen atoms have a spectral line of 21cm wavelength (1420 MHz in frequency) at rest, and as the Universe expands, it is redshifted to lower frequencies given by $\nu=1420/(1+z)$ MHz, where $z$ is called the redshift of the emitter. The Tianlai array is currently observing in the 700-800 MHz band, corresponding to redshifts $z = 0.78$-$1.03$ for the neutral hydrogen line.

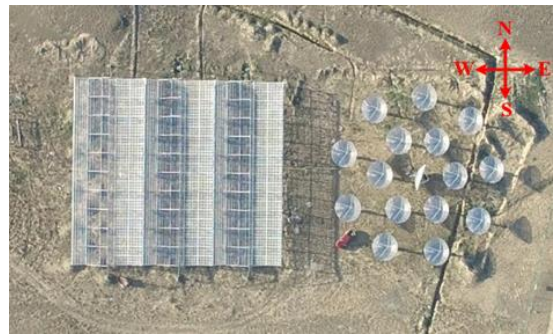

**Figure 1** Photograph of the Tianlai array. Currently the Tianlai project includes an array of 6 m dishes (Tianlai dish array) and an array made up of three, 15 m wide and 40 m long parabolic cylinder antennas with a total of 96 dual polarization feed units (Tianlai cylinder array). These arrays are located in a radio quiet site in Xinjiang, China (longitude 91°48´20˝ E, latitude 44°09´08˝ N).

To obtain accurate cosmological observations, the beam pattern of the telescope needs to be measured and calibrated. Due to the geographic condition of the site, it is not practical to put an artificial source at a suitable height near the array for calibration. Also, we have not found a suitable satellite transmitting in the 700-800 MHz band. Among astronomical sources with known position and flux density, Cassiopeia A (Cas A) is one of the brightest. We measured the beam pattern of a single dish telescope using Cas A. The result shows that only the main lobe is measurable, for although Cas A is the brightest astronomical source in this band, it is not strong enough to permit measurements of the side lobes.

As a demonstration of the principle, we calibrated a single dish telescope in the Tianlai project using a micro UAV. The signal transmitting system, the flight path of the UAV, the setup of the experiment and the data processing procedure are described below in detail. We compare the UAV-based measurement both with the traditional method of using an astronomical source (Cas A) and with simulations.

## 2 Description of measurement system

Astronomical sources are often used as calibrators for measuring the beam of radio telescopes. However, the available signal strength is limited for the beam measurements of some telescopes. The UAV carrying an emitter can fly along a planned path as a calibrator source for this purpose. A schematic of the measurement is shown in **Figure 2**. The measurement signal is transmitted from the UAV system, which flies at the specified height to scan the AUT. The signal is collected by the receiver of the dish telescope system. Combining knowledge of the position of the UAV and the received power, one can map the beam pattern after data processing. The measurement precision of the position of the UAV is at the centimeter level (1-2 cm).

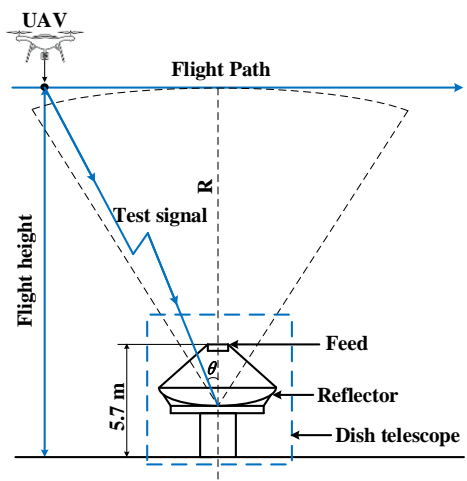

**Figure 2** Schematic of the beam measurement. The AUT is a dish antenna in this experiment. The UAV equipped with the calibration signal transmitting system flies along a specified flight path so that the azimuth and the elevation of the AUT remains unchanged during measurement. The flight height should be beyond the far field distance of the AUT.

Some limitations of this approach must be kept in mind. Both the battery capacity and the cargo space under the UAV are limited, so the transmitting antenna should be light and convenient to mount. The hovering time of most small civilian UAVs are a few tens of minutes (in our case 30 minutes), and the flight height is a few hundred meters (in our case 500 meters), limited by the capability of the UAV as well as the civil aviation safety regulations. This height may not fulfill the far field distance requirement for beam measurement for some antennas, but it is adequate for the Tianlai dishes.

### 2.1 UAV

The UAV used in this experiment is a DJI Matrice 600 PRO. Some of the characteristics of this model are listed in **Table 1**. The main features relevant for this experiment are as follows:

1. The signal transmitting system is mounted directly under the UAV via two carbon fiber extension rods. The polarization direction of the transmitting antenna is parallel to the ground during the measurement.

2. Flight operations such as taking off, hovering at waypoints, and landing are programmed in advance and automatically carried out by the UAV.

3. With the built-in GPS system, the positioning precision of the UAV is about 1 meter, which would lead to a 1° error in the beam measurement.

**Table 1** Characteristics of the UAV.

| DJI Matrice 600 PRO | |
| --- | --- |
| length × width × height | 1.8 m × 1.6 m × 0.8 m |
| Max payload | 6 kg |
| Hovering Time | 38 minutes |
| Flight height (state control) | 500 m (max) |
| Hovering accuracy (with GPS) | Horizontal: ± 1.5 m  Vertical: ± 0.5 m |
| Hovering accuracy (with D-RTK system) | Horizontal: ± 1 cm  Vertical: ± 2 cm |

In order to improve the positioning accuracy, Virone et al. [10] used a motorized total station theodolite to measure the real-time position of the UAV, and achieved centimeter level accuracy in their experiment. However, the operation of the total station theodolite is complicated. In this work, we used the DJI- real time kinematic (D-RTK) difference system to obtain GPS coordinates of the UAV. The RTK system enhances the precision of the GPS positioning by re-broadcasting the phase of the signal carrier that it receives; the mobile unit on the UAV then can determine its position relative to the base station. The D-RTK system is divided into the ground unit and the airborne unit, as shown in **Figure 3**. The ground unit is placed at a precisely measured geodetic datum point near the AUT before each flight. The transmission range of the ground unit is 600 m with a total power of 10 dBmW. Using the D-RTK system, the relative positioning precision is improved to the centimeter level[1].

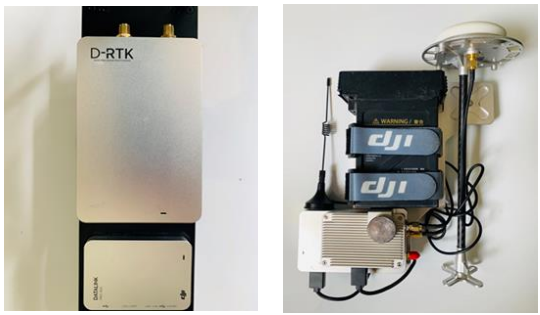

**Figure 3** The D-RTK airborne unit (left) and ground unit (right). The airborne unit is mounted on top of the UAV. An omnidirectional antenna (colored black in the photo) is used by the D-RTK ground system to transmit the re-broadcasted GPS signal.

## 2.2 Calibration signal transmitting system

The system that transmits the artificial calibration signal consists of a broad-band noise source, band pass filter, attenuator, battery, transmitting antenna, and a shielding box. The voltage stabilizer, the band pass filter and the noise source are placed inside the shielding box. **Figure 4** shows the UAV equipped with the signal transmitting system. The rated voltage of the noise source is 12 V. In order to ensure stable output of the noise source, we used a voltage regulator to regulate the 16.8 V battery voltage at 12 V. The output frequency range of the noise source is 0.1 MHz -1.5 GHz and the total power is about 17 dBmW, as shown in **Figure 5**. The pass band of the filter is 700-800 MHz, which is consistent with the observation band of the receiving system. To reduce the weight, the shielding box is made of plastic, with its external surface covered with copper foil. The total weight of the signal transmitting system and the D-RTK air system is 1.3 kg. The hovering time of the UAV equipped with these devices is over 30 minutes.

The transmitted power needs to be adjustable to measure the different parts of the AUT's beam without saturating the dish telescope's receiver. Combinations of attenuators provide attenuations ranging from 3 dB to 43 dB in the signal transmitting system to produce signals within the dynamic range of the receiver.

A dipole suitable for the measurement range of 700-800 MHz is selected as the transmitting antenna, as shown in **Figure 6**. The transmitting antenna is mounted below the shielding box and is oriented horizontally. Its direction can be adjusted to be either parallel to or perpendicular to the nose of the UAV.

The vertical distance from the GNSS (Global Navigation Satellite System) antenna of the D-RTK air system to the transmitting antenna is 0.5 m.

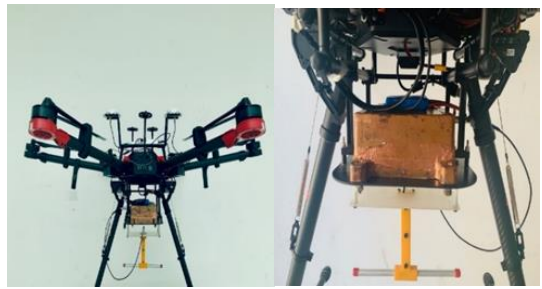

**Figure 4** Photograph of the UAV equipped with the calibration signal transmitting system. The D-RTK airborne unit is mounted on top of the UAV, and the dipole antenna that transmits the calibration signal is mounted below the UAV, parallel to the ground.

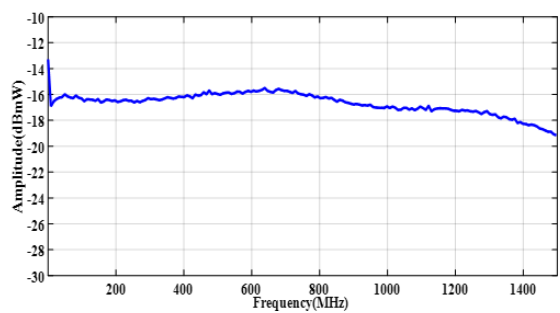

**Figure 5** Emitted power of the noise source. The emitted power is measured by a spectrometer with a resolution bandwidth (RBW) of 1 MHz. The output power is about -16 dBmW in the observation band of 700-800 MHz.

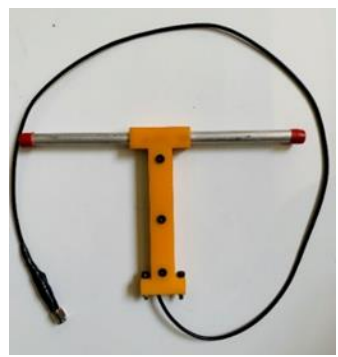

**Figure 6** Photograph of the transmitting dipole antenna. It is an aluminum tube with an outer diameter of 8 mm and length of 192 mm. The designed central frequency is 750 MHz

The beam pattern of the transmitting antenna in the frequency range of 700-800 MHz is measured first with the shielding box and then again when the dipole and shielding box are mounted on the UAV. **Figure 7** shows a comparison between the smoothed measurement data and an electromagnetic simulation of the transmitting dipole antenna by itself at 730 MHz. In our flights the transmitting dipole antenna is installed either parallel to or perpendicular to the nose of the UAV (i.e. the direction of the flight path). In the parallel case, the E-plane of the transmitting dipole affects the measurement, while in the perpendicular case, the H-plane affects the measurement. So, in Figure 7 we plot the beam profile in the E-plane for the parallel case; and H-plane for the perpendicular case. In either case, with the transmitting dipole located below the UAV, the downward direction is defined as 0° and the direction of the UAV nose is defined as -90°. In our experiment, the maximum angle between transmitting antenna and the AUT is about 23°, so here only beam angles of up to +/-25° are compared.

---

[1] https://www.dji.com/cn/datalink-pro/info#specs

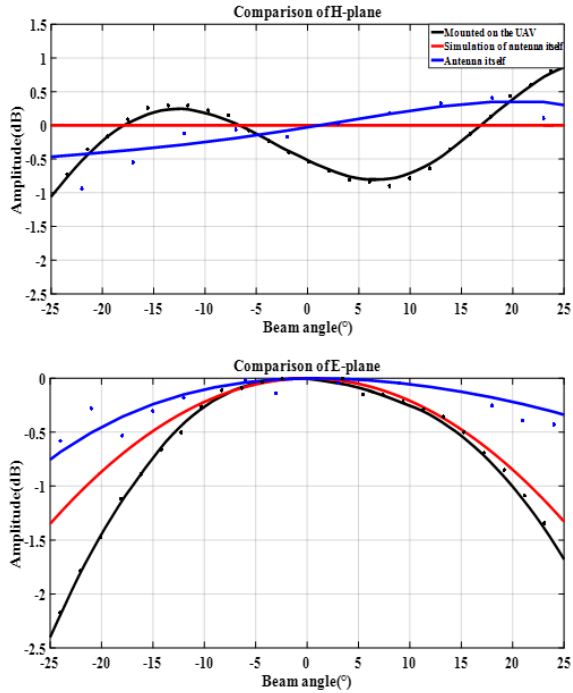

**Figure 7** Comparison of the simulated and measured beam patterns of the transmitting dipole antenna at 730 MHz. The top panel shows the H-plane for the case of transmitting dipole perpendicular to the flight direction, the bottom panel shows the E-plane for the case of transmitting dipole parallel to the flight direction. The red curve is the simulated pattern of the dipole by itself, the blue and black curves are the smoothed pattern from measurements with only the shielding box mounted, and both shielding box and UAV mounted, respectively. In data processing, we correct the AUT's beam according to the measured beam of the UAV-equipped transmitting antenna.

Note that these measurements of the transmitting dipole are taken on the ground, so the difference between the measurements and the simulation is not entirely due to the effect of the attached shielding box and the UAV, but may have some contributions due to ground reflections. Nevertheless, the difference between either of the two measurements and the simulation is not large, the biggest difference within the range is less than 0.5 dB for the E-plane and 1 dB for the H-plane, and near the center the difference is even smaller than this. Below, we use the measured, smoothed UAV-mounted dipole beam for correction. As shown in Figure 7 the beam profiles of the transmitting dipole are slightly asymmetric, and in our measurement of the AUT we also found slight asymmetry in the raw data at the level of 1-2 dB. Applying the correction reduces the asymmetry in the measured beam of the AUT.

### 2.3 Receiver of the dish telescope

The calibration signal is collected by the receiver system of the dish telescope. After amplification by the low noise amplifier (LNA), the received signals are transmitted to the telescope correlator through optical fibers. Each dish telescope receives two signals through a dual-dipole feed (described in section 3.1). There are 512 frequency channels currently in the receiver system and each channel covers about 244 kHz, corresponding to the frequency band of 685-810 MHz [11]. The integration time is 1 s and the dynamic range is about 30 dB.

## 3 BEAM CALIBRATION OF DISH TELESCOPE

### 3.1 The 6 m dish telescope

A Tianlai dish telescope is shown in **Figure 8**. It has a diameter of 6 meters, with metal mesh reflector and primary focus feed. The antenna has an altazimuth mount. During the measurement, the telescope is pointed to 180° azimuth (due south) and 88.5° in elevation (near the zenith). This high elevation angle is chosen to reduce the ground pickup as much as possible. The dual linear polarization feed of the dish is of a crossed-dipoles type, with one of the dipoles (H) oriented horizontally (parallel to the elevation axis of the telescope, along the E-W direction), and the other (V) is oriented along the N-S direction.

For this dish, in the observation band of 700-800 MHz, the minimum far field distance is (taking the frequency of 800 MHz)

$$d = \frac{2D^2}{\lambda} = 192 \text{m} \qquad (1)$$

Where $D$ is the antenna aperture and $\lambda$ is the wavelength.

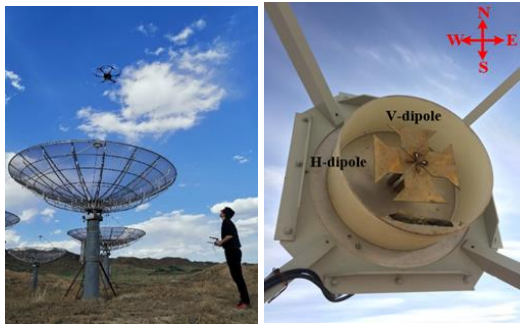

**Figure 8** A single 6 m dish telescope (left) and the feed used in the telescope (right). The feed is a dual-dipole antenna supported by four metal struts. The two dipoles of the feed are perpendicular to each other, and are oriented in the E-W direction and the N-S direction respectively when the telescope is pointing vertically up towards the zenith. We label these two dipole as H-dipole (horizontal) and V-dipole (vertical).

### 3.2 Flight path planning

The positions of the Tianlai dish array antennas are shown in **Figure 9**; 15 dishes are distributed in two concentric circles around the center dish. In our experiment, the No. 6 dish telescope located at the southern tip of the close-packed array is chosen to be the AUT, so that when the UAV is flying above and slightly to the south of the array, it is least affected by the other antennas.

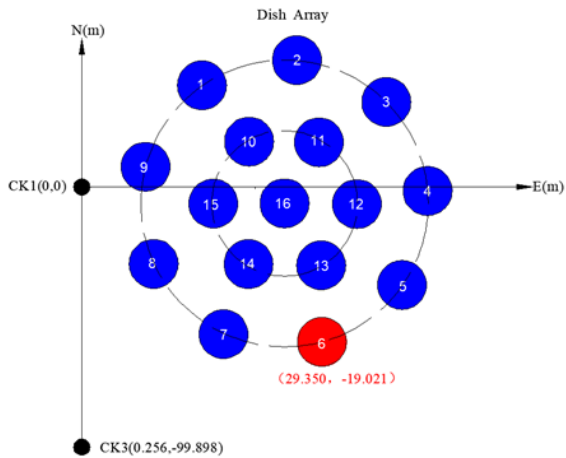

**Figure 9** Positions of the antennas in the dish array. Relative coordinates between each dish and the origin were measured accurately by a total station theodolite.

Two pre-determined geodetic points CK1 and CK3 are also shown on the map. The D-RTK ground unit is placed at the geodetic datum point CK3. By using the GPS coordinates in the D-RTK system and the relative coordinates between CK3 and the AUT, the distance and direction of the UAV with respect to the AUT can be calculated.

The UAV has two flight modes available: (1) the continuous flight mode in which it flies along a given path with uniform speed; (2) the hovering flight mode, in which the UAV flies along the path to a series of specific waypoints and hovers for a few seconds at each. The hovering time can be adjusted between 1 s to 30 s. Since the integration time for the Tianlai dish array is about 1 s, it is difficult to use the continuous flight mode. Therefore, we use the hovering mode in our experiment. During the tests, the UAV requires about 2 s to stabilize at each waypoint, so we set the hovering time to 5 s in the experiment.

The UAV was flown along the east-west and the north-south directions, forming a cross centered above the AUT position as shown in **Figure 10**. For each of these two flight paths, the measurement is taken with the polarization of the emitter both parallel to and perpendicular to the flight direction. All measurements were made when the local weather was a gentle breeze to avoid deviations induced by windy conditions. As discussed above, the distance to the AUT should exceed 192 m to satisfy the far-field condition. During the experiment we have made measurements with two heights: 300 m and 480 m above the ground (the operation is restricted to below 500 m above the ground level per regulation). Note the altitude of the ground is about 1500 m above sea level. The polarization orientation of the dipole is set to be parallel (perpendicular) to the flight direction in the E-plane (H-plane) measurement. Because the transmission distance of the D-RTK ground system is limited, we set the length of each flight path to be 400 m. The spacing between adjacent waypoints is 5 m so that each flight path has a total of 81 waypoints.

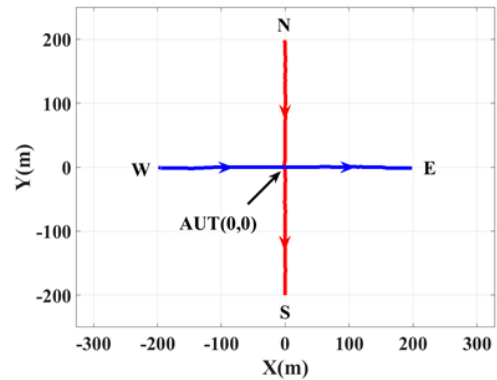

Figure 10 The flight path during the experiment. The flight paths were along lines in N-S direction and E-W direction. The AUT is at the origin of the coordinate system. The arrows show the flight direction of the UAV.

# 4 RESULTS

## 4.1 Measurement of the UAV noise

The signals collected by the dish receiver include the transmitted calibration signal, noise, and possibly radio frequency interference (RFI) from electronics on the UAV itself. The operating frequency of the UAV controller signal is 2.4 GHz, which is far outside the 700-800 MHz bandpass of the receiver. The transmitting frequency between the D-RTK ground system and the air system is 430-432 MHz, much closer to the receiver bandpass. RFI from the D-RTK system and other electronics on the UAV are assumed to generate a constant `noise' level. This noise can be measured easily during the experiment by flying the UAV along the designed path with the calibration source turned off. The measured noise level during the flights of the UAV is shown in **Figure 11**. The measured noise is slightly higher (at a percent level) when the UAV is over flying the main lobe of the AUT. To compensate for this, we subtract this UAV-induced noise in later data processing.

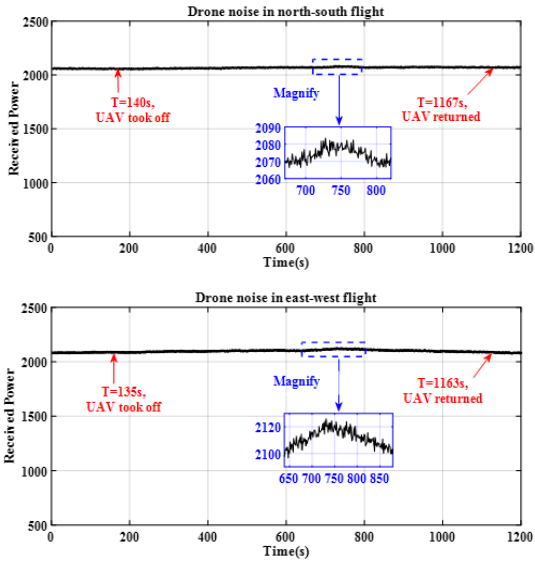

**Figure 11** Measurement of the noise during flights of the UAV. The units of the vertical axis are arbitrary. The noise has a tiny peak when the UAV is flying above the telescope.

## 4.2 Data processing

The polarization of the transmitting antenna is oriented either parallel or perpendicular to the polarizations of the dual-polarization AUT feeds during the measurements. The received signal from one polarization of one feed (copolar response) is shown in **Figure 12**. The main lobe and side lobes are evident in the blue curve. There is also a noise floor of about 2000 in the readout; when this is subtracted the variance of the residual of the noise floor is about 3, much smaller than the signal level.

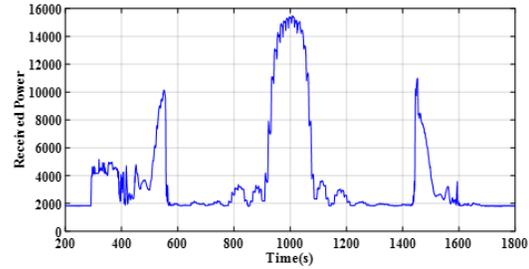

**Figure 12** The received signal of one polarization of the dual-polarization feed during one flight. The units of vertical axis are arbitrary and the number is not equal to the actual received power. The center frequency is 730 MHz with 5 MHz bandwidth. The feed polarization is parallel to the transmitting antenna. The received power between 290-570 s and between 1430-1750 s corresponds to the ascent and descent of the UAV.

The time interval for saving the GPS data is set to be the same as the integration time of the receiver. The GPS coordinates of the UAV location are stored in the UAV controller along with time stamps at a rate of once per second. The received signals are measured by the correlator of the telescope.

Because of the finite dynamic range of the dish receiver, the beam was measured many times with different attenuators.

Taking the measurement in the north-south direction at 730 MHz with bandwidth of 5 MHz as an example, the data processing procedure is shown in **Figure 13**. The steps are as follows:

1. Match the coordinates of the UAV location to the signal recorded by the receiver at every second during the experiment, and select the 81 test waypoints in the data (the blue dots shown in the figure).

2. Use the UAV GPS data to derive the relative positions between the UAV and the AUT and calculate the beam angles.

3. Subtract the background noise and the UAV-induced noise.

4. Compute the beam pattern by normalizing the received power at each waypoint with the expectation according to the inverse-square law. Finally, correct the measured beam pattern according to the measured beam of the transmitting antenna (when mounted on the UAV) at each waypoint.

We have measured the copolar and crosspolar response in both the E-plane and H-plane of the H and V polarizations of the AUT feed. We have used these responses to calculate the cross polarization isolation. If the copolar polarization is denoted X and crosspolar polarization denoted Y, we could measure the crosspolar response by auto-correlation <YY>/<XX> or cross-correlation (<XY>/<XX>)$^2$, these are given in **Table 2.** Both fractions are at the percent level. Part of this may be due to polarization-leakage, but part of it may be caused by the misalignment of the transmitter and AUT feed. In principle these two effects may be separated by fine-tuning the orientation of the transmitter antenna, until a minimum is found in the auto-correlation of the orthogonal polarization, and taking the cross-polarization measurement at that point. In practice this approach is difficult, as it is difficult to make precise adjustment for the orientation of the UAV, and the UAV position also fluctuates during the measurement.

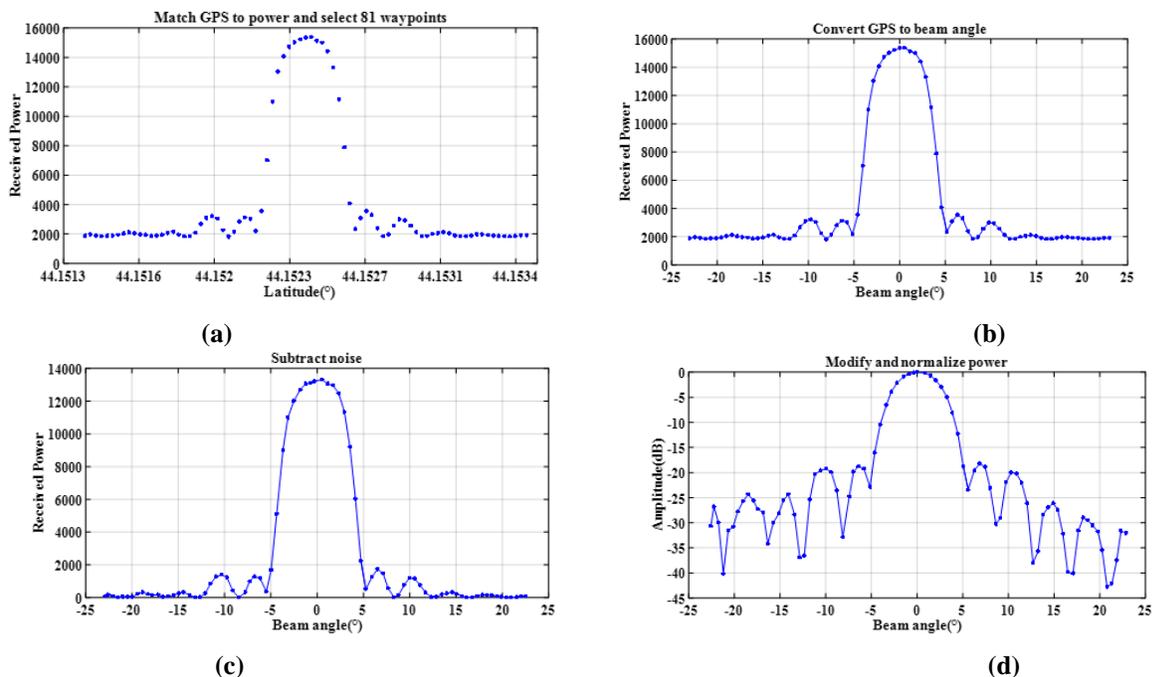

**Figure 13** Data processing procedure. Pictures from (a) to (d) are corresponding to the four data processing steps respectively. The units of vertical axis are arbitrary.

**Table 2** Cross polarization isolation

| Frequency | Auto correlation | Cross correlation |
|---|---|---|
| 730 MHz | 1.17% | 1.06% |
| 750 MHz | 1.76% | 1.65% |
| 770 MHz | 2.31% | 0.925% |

## 4.3 Verification with repeated measurements

We have performed the measurements a few times to check the reliability of the method as well as the stability of the system. The H-plane beams were first measured during August and October 2018. Then in April 2019, we measured the E-plane, and also adopted a different UAV flight height to verify the previous results. Taking the H-plane of the dipole in the north-south direction as an example, the comparison between different times and different flight heights is shown in **Figure 14**. The shape of the main lobe of these three plots agree well with each other. The measured half power beam width (HPBW) are 5.17, 5.10, and 5.07 degrees respectively, very close to the design specification of the Tianlai dish telescope of 5 degrees. The positions and widths of the side lobes of the different measurements are quite consistent with each other. The good agreement between the different measurements shows that this UAV-based measurement is stable and the results are reliable.

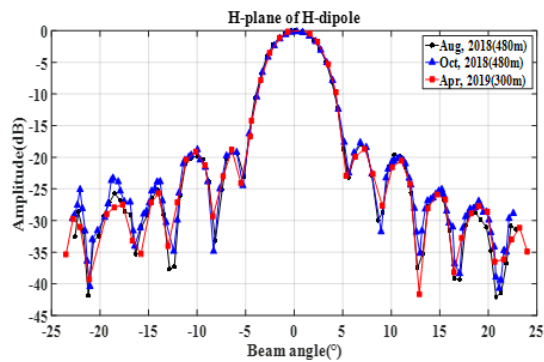

**Figure 14** Comparison of the H-plane patterns. The measurement center frequency is 730 MHz with 5 MHz bandwidth. The flight direction is the north-south direction. The flight height is 480 m in August and October 2018. In April 2019, we set the flight height to be 300 m.

We mapped the beam patterns at 730 MHz, 750 MHz, and 770 MHz. Results of the measurement are shown in **Figure 15**. We can see there are indeed some small differences in the main lobe, with the beam width slightly smaller for the higher frequencies, as expected. The difference are larger in the side lobes.

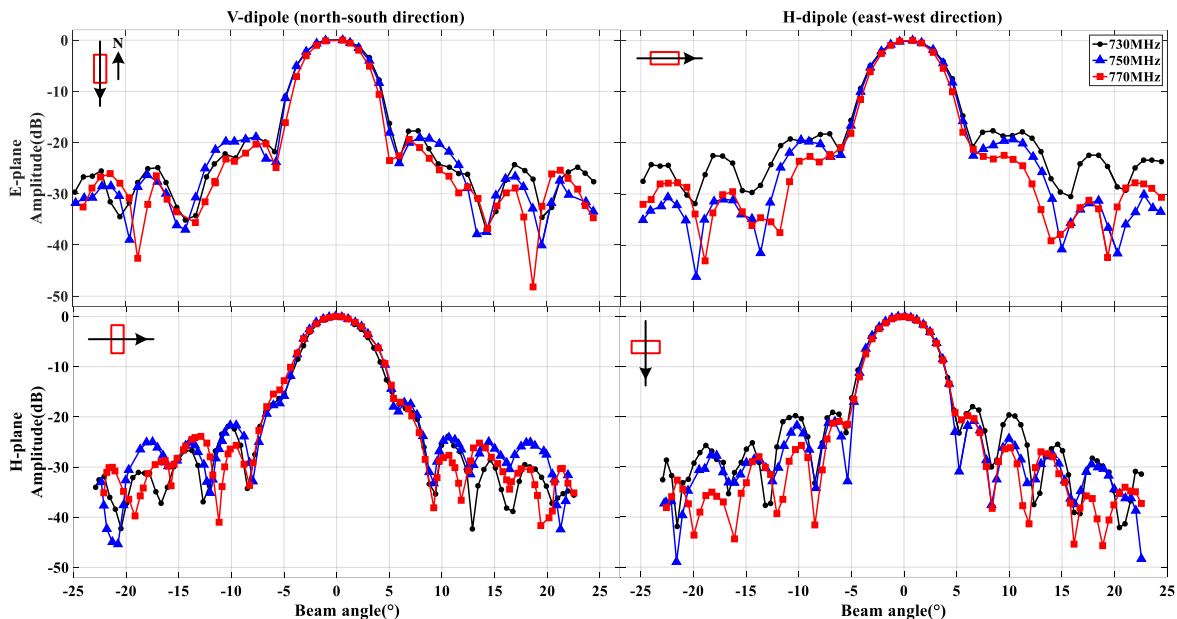

**Figure 15** Beam pattern of dish telescope No.6. We measured the beam pattern at 730 MHz, 750 MHz and 770 MHz. The bandwidth at each frequency is 5 MHz. The E-plane and H-plane of each dipole in the dish feed are measured respectively. In the top left corner of each beam pattern, the red rectangle shows the polarization orientation of the transmitting antenna (along the long axis of the rectangle), the black line is the UAV flight path, and the arrow shows the flight direction.

## *4.4 Comparison with calibration using astronomical source*

We also compare our UAV measurement with calibration using the strong astronomical radio source Cas A. For this measurement we calculated the expected transit position of Cas A in advance, pointed the AUT to a position where the source was expected to pass through (in this measurement, it is elevation 48.50°, azimuth 45.27°) and the receiver was set to record data. To avoid RFI, the motion control system of dish telescope were turned off and the azimuth and the elevation remained unchanged during the measurement.

The comparison of these two methods is shown in **Figure 16**. The red dashed curve in the figure is from the Cas A measurement, and the blue dotted curve is the result of the UAV measurement. The black curves are the discrepancies of the main lobe; the maximum discrepancy is less than 3 dB. Using Cas A as a source we could only measure the beam over an angular range of about 10°. The figure shows that the signal is not strong enough to measure the side lobes. For the UAV measurement, the range is extended out to the fourth side lobe. The discrepancies of HPBW between the UAV and the Cas A measurement are 0.33° and 0.54° for the E and H plane respectively. It should be noted that during the Cas A transit, the track of the Cas A is an arc, so it is not exactly on the E-plane or H-plane of the two polarizations, but the difference is small. Despite this, the beam pattern measured by the UAV shows good agreement in the main lobe with the Cas A measurement. This shows that the UAV-based system is feasible for calibrating the dish telescope.

## *4.5 Comparison with simulation*

The beam pattern of the telescope is also simulated using the CST Microwave Studio electromagnetic software. The model is shown in **Figure 17**. We simulated the beam of the AUT both as an isolated dish telescope, and in the presence of multiple dishes. In the latter case, we added four nearby dishes (No.5, No.7, No.13, and No.14) in the model. The comparison between the simulation and the UAV measurement is shown in **Figure 18**. The measured position and width of each lobe shows good agreement with the simulation. Compared with the isolated dish case (black curve), when multiple dishes are included in the model (red curve), the beam pattern becomes asymmetric with respect to the beam angle, thanks to the influence of the nearby dishes. Furthermore, a "shoulder" appears at ±6°, which agrees better with the measurement (blue curve). There is an obvious difference of amplitude in the side lobes between the simulation and measurement, which is perhaps due to the difference between the ideal 3D model used for simulation and the actual specimen. The elevation of the measured side lobes may raise from the imperfection in the reflector, as well as phase errors caused by the scattering by the ground and surroundings, and noise in the receiver system.

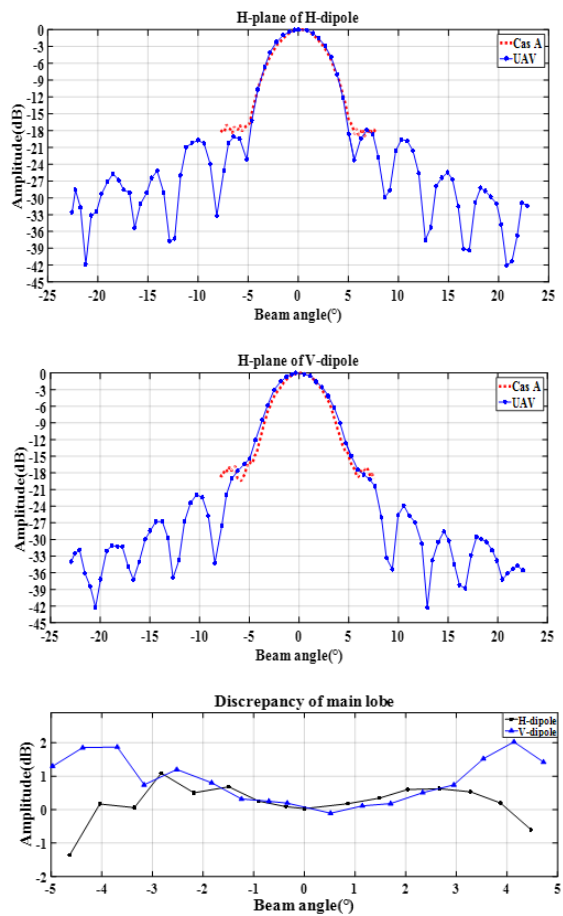

Figure 16 Comparison of beam patterns measured by the UAV and Cas A at 730 MHz with 5 MHz bandwidth. The blue curves are the H-plane pattern of two dipoles in the AUT feed. The bottom figure shows the difference (UAV-Cas A) of the main lobe in -5°-5°. In the measurement by Cas A, there are no obvious side lobes but only one peak, which represents the main lobe.

### 4.6 Sources of Error

1. Position errors of the UAV. We checked the deviations of the UAV during the measurements. The vertical positioning accuracy of the D-RTK is 2 cm, which corresponds to an error of signal intensity of less than 0.01%. The horizontal deviations between the actual path of the UAV and the theoretical path were within 2 m, induced by a gentle breeze, and caused an angular error of less than 0.38 ° with respect to the E-plane (H-plane). Note that the UAV actual position was measured by the D-RTK system and is accurate at the centimeter level.

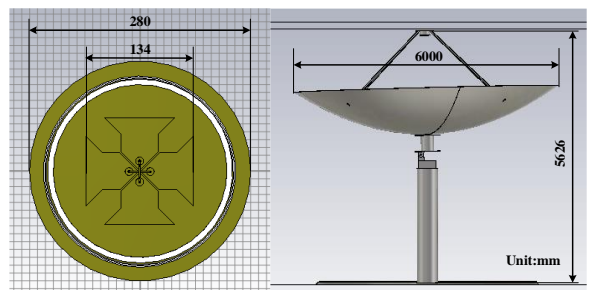

Figure 17 The model of the feed (left) and the dish telescope (right). The model size is built on a 1:1 scale with the telescope. The feed is supported by four metal struts, the angle between each strut and the dipole is 45°. The rotating mechanical structure under the reflection panel is simplified, and the strut is modelled by a simple cylinder. To simplify the simulation of the complex model, the material was set to be perfectly electric conducting (PEC) and the ideal electromagnetic environment was used in the simulation.

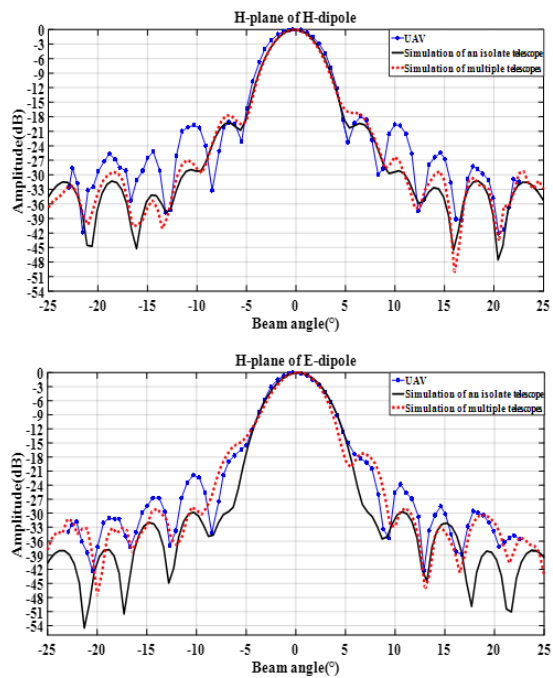

Figure 18 H-plane pattern comparison between measurement and simulations. The curves show good agreement on the position and width of lobes. The difference in amplitude is presumably caused by environmental factors not included in the simulation, such as scattering from the ground or other antennas in the array.

2. The position reported by the GPS/D-RTK system corresponds to the phase center of the GPS receivers. The position of the calibration signal source on the UAV is the phase center of the transmitter antenna on the UAV, and the reflection off the UAV may further shift its effective position. These differences in the position may incur differences of up to 10 cm, which will not be the dominant source of error.

3. Polarization of the transmitting antenna. The

polarization of the dipole can deviate from the north-south (east-west) direction due to the yaw of the UAV. We verified that the deviation of the yaw was stable and within 2 ° during the measurement. The error caused to the copolar measurements by polarization mismatch was within 3.5%.

4. Uncertainty in the beam of the transmitter on the UAV. The beam of the transmitting antenna is altered by reflections from the UAV itself. We used the measured beam pattern of the transmitting dipole antenna mounted on the UAV to correct the measured beam pattern of the AUT, but there are measurement errors in this pattern. The maximum difference between the measured beam pattern and the simulated ideal dipole is less than 1 dB in the range of our measurement, and the difference between the actual pattern and our measurement should be even less than this number. Conservatively, we estimate the largest error induced by this correction is less than 1 dB. This is perhaps the primary source of error in our measurement.

In addition, some of the signal transmitted from the dipole was also reflected by hills near the AUT, the other dishes in the array, and the ground. This scattered signal may cause phase errors and raise the level of the side lobes. Imperfections in the reflector may result in an asymmetric beam pattern.

## 5 Discussion

This paper describes the beam measurement of one of the single dish radio telescopes in the Tianlai dish array using an UAV. Our results for different times and flight heights demonstrate that the UAV-based method is stable and reliable. The beam pattern out to the fourth side lobe is mapped by the UAV, while with the astronomical source Cas A only the main lobe can be measured for this dish. The beam patterns measured using the UAV and Cas A show good agreement in the main lobe. The position and width of side lobes are also in good agreement with the result of an electromagnetic field simulation. The difference of amplitude is probably caused by the actual electromagnetic environment of the observatory. The inexpensive UAV-based approach is flexible, and the measurement range is wider than with an astronomical source.

This approach can also be applied to the calibration of other radio telescopes. A limitation here is due to the far field condition for the measurement. From Equation (1), the diameter of the AUT is limited to

$$D \leq \sqrt{\lambda d / 2} \qquad (2)$$

If the UAV is flying directly above the AUT, this distance is limited by the height of the UAV. For example, in our case, the height limit is 500 m, so at 800 MHz, the maximum diameter of the AUT allowed is 9.65m. Alternatively, for an antenna of 6 m diameter, this translates to a limit on the shortest measurable wavelength of 14.4 cm, or highest frequency of 2.08 GHz. However, this limit can be lifted if the AUT is not required to point at the zenith, but can be tilted to a lower elevation. In that case, the far field requirement can be fulfilled if the UAV is flying at a certain distance away, though in such a case the ground pickup may cause some error.

## Acknowledgments

This experiment has been funded by the National Natural Science Foundation of China (NSFC) project entitled "Research on near and far-field measurement technology of large and medium-sized radio telescope based on a micro unmanned rotorcraft" (U1631118), the NSFC key project "Theoretical and Observational Studies on Neutral Hydrogen" (11633004), the CAS Frontier Science Project "Cosmological Neutral Hydrogen Survey and Data Analysis" (QYZDJ-SSW-SLH017), and The CAS Interdisciplinary Innovation Team(JCTD-2019-05). Work at UW-Madison is partially supported by NSF Award AST-1616554. We thank Li Deng, Bin Yue and Weiyang Wang for assistance in measuring the drone antenna pattern. We also thank Xia Chen and ZhiGuang Lin from the DJI Company for assistance in controlling and using the drone.

## References

1. A. R. Neben, R. F. Bradley, and J. N. Hewitt, et al. "Measuring phased‐array antenna beam patterns with high dynamic range for the Murchison Widefield Array using 137 MHz ORBCOMM satellites," Radio Science, vol. 50, no. 7, pp. 614–629, 2015. (journal)

2. J. Baars, "The measurement of large antennas with cosmic radio sources," IEEE Transactions on Antennas & Propagation, vol. 21, no. 4, pp. 461–474, 2003. (journal)

3. C. Chang, C. Monstein, A. Refregier, et al. "Beam Calibration of Radio Telescopes with UAVs," Publications of the Astronomical Society of the Pacific, vol. 127, no. 957, pp. 1131–1143, 2015. (journal)

4. A. M. Picar, C. Marque, M. Anciaux, et al. "Antenna pattern calibration of radio telescopes using an UAV-based device," IEEE International Conference on Electromagnetics in Advanced Applications, 2015. (conference proceedings)

5. D. C. Jacobs, J. Burba, J. Bowman, et al. "The External Calibrator for Hydrogen Observatories," IEEE Conference on

Antenna Measurements & Applications, 2016. (conference proceedings)

6. E. D. L. Acedo, P. Bolli, F. Paonessa, et al. "SKA aperture array verification system: electromagnetic modeling and beam pattern measurements using a micro UAV," Experimental Astronomy, vol. 127, pp. 1–20, 2018 (journal)

7. X.L. Chen, "Radio detection of dark energy——the Tianlai project," Scientia Sinica, vol.5304, no. 36, pp. 43–56, 2011 (journal)

8. Y. Xu, X. Wang, X. Chen, "Forecasts on the Dark Energy and Primordial Non-Gaussianity Observations with the Tianlai Cylinder Array, " Astrophysical Journal, vol.798, no. 1, pp. 40, 2015. (journal)

9. J. Zhang, R. Ansari, X. Chen, et al. "Sky reconstruction from transit visibilities: PAON-4 and Tianlai Dish Array, " Monthly Notices of the Royal Astronomical Society, vol. 461, no. 2, pp.1950–1966, 2016. (journal)

10. G. Virone, A. M. Lingua, M. Piras, et al. "Antenna Pattern Verification System Based on a Micro Unmanned Aerial Vehicle (UAV) , " IEEE Antennas and Wireless Propagation Letters, vol. 13, pp.169–172, 2014. (journal)

11. S. Das, A. J. Cianciara, C. J. Anderson, et al. "Progress in the construction and testing of the Tianlai radio interferometers," Proc. SPIE 10708, Millimeter, Submillimeter, and Far-Infrared Detectors and Instrumentation for Astronomy IX, 2018. (journal)